# Regenerative Soot-III: Role of the kinetic and potential sputtering in the regeneration of the soot

Shoaib Ahmad[1,2,*] and M. N. Akhtar[1]

[1]*Accelerator Laboratory, PINSTECH, P. O. Nilore, Islamabad, Pakistan*

[2]*National Centre for Physics, Quaid-i-Azam University Campus, Shahdara Valley, Islamabad, 44000, Pakistan*

[*]Email: sahmad.ncp@gmail.com

## Abstract

We have used the photoemission spectroscopy of the graphite hollow cathode sooting discharge to identify the roles played by the kinetic and potential sputtering, respectively. Our indicators are the relative densities of the sputtered carbon C, the Ne metastable atoms, and the ionized components ($Ne^+$) of the regenerative sooting plasma. We find that the metastable atoms are the main agents for the regeneration of the soot.

Soot formation in carbon vapor and arc discharges and also in the laser ablated graphite plasmas leads to carbon clusters $C_m (m>2)$ and the fullerenes [1–4]. In our case a graphite hollow cathode produces regenerative soot in Ne discharges [5]. Sooting plasma implies an environment that is conducive to the continuous process of the graphite cathode wall sputtering. This introduces monatomic carbon $C_1$ and higher clusters $C_m (m>2)$ into the plasma and that further leads to the wall coverage with neutral, excited, and positively charged $C_m^{0,*,+}$ species. Once the discharge has been initiated, its prolonged operation ensures the coverage of the cathode walls with a sooting layer.

This layer, in turn provides a sooting character to the initially pure Ne plasma. The ionized and metastable atoms are the agents for the cathode wall sputtering that gives the plasma its carbon content. The wall erosion and the soot formation are closely linked with the plasma properties. One significant indicator of clustering to have taken place in the hollow cathode discharges is to monitor the emitted species by mass spectrometry. Using a velocity analyzer, we reported on the formation and emission of carbon clusters from a very wide range of C clusters from $C_2$ to $C_{4530}$ [6]. Recently, another study on the photoemission spectroscopy and mass analysis of the atomic and cluster species sputtered from heavy ion irradiated graphite surface was conducted [7]. The aim was to identify the role of the kinetic sputtering of graphite in a non-regenerative mode. Negative atoms and clusters $C_1^-$, $C_2^-$, and an order of magnitude smaller intensities of $C_3^-$ and $C_4^-$ were detected in the mass spectra.

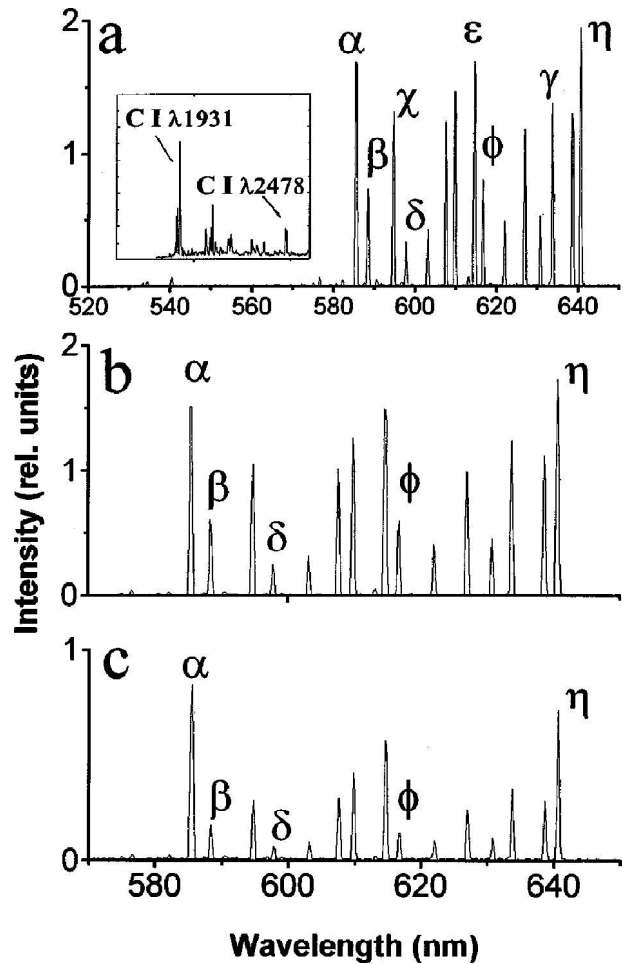

FIG. 1. Photoemission spectra of the hollow cathode cluster source are shown for the emission lines of Ne I. (a) is plotted for $i_{dis}$=150 mA, also shown in the inset is the ×10 enlarged spectrum between 180 and 260 nm where the CI's main emission lines $\lambda$ =1931 Å and $\lambda$ =2478 Å, can be seen. Also shown are the emission lines labelled $\alpha(\lambda =5852Å)$, $\beta(\lambda =5882Å)$, $\chi(\lambda =5945Å)$, $\delta(\lambda =6030Å)$, $\varepsilon(\lambda =6145Å)$, $\phi(\lambda =6164Å)$, $\gamma (\lambda =6334Å)$, and $\eta (\lambda =6404Å)$. All these lines decay to the 3s levels, therefore, producing the Ne metastables. (b) and (c) show the same lines at $i_{dis}$=100 and 50 mA, respectively.

The photoemission spectra of the irradiated surface, however, revealed the sputtering of $C_1$, $C_1^+$, and also $C_2$. In addition, the relative contributions of the various C emission lines showed a large percentage of metastable C atoms among the sputtered species.

In this letter, we report the results on the photoemission spectra from a cusp field, hollow cathode, graphite ion source that was described in detail elsewhere [5]. Photoemission spectroscopy of the regenerative soot is employed to reveal its underlying processes and the mechanisms. The data are obtained as a function of the discharge current $i_{dis}$ and the pressure $P_g$ of the source gas Ne. The excited and ionized $C^{*,+}$ reach and interact with the cathode just like the excited and charged neon $Ne^*$ and $Ne^+$. We have attempted to identify the two complementary mechanisms of cathode erosion, i.e., the kinetic and potential sputtering. Figure 1 shows three spectra from the hollow cathode source operating at Ne pressure $P_g$= 0.1 mbar at discharge current $i_{dis}$= 150, 100, and 50 mA, respectively. Figure 1(a) at $i_{dis}$= 150 mA and $V_{dis}$ = 600V shows the excited neon atomic ($Ne^*$) emission lines denoted with the spectroscopic notation-NeI in the 5800 and 6500Å range. The lines are labelled as $\alpha(\lambda = 5852Å)$, $\beta(\lambda = 5882Å)$, $\chi(\lambda = 5945Å)$, $\delta(\lambda = 6030Å)$, $\varepsilon(\lambda = 6145Å)$, $\phi(\lambda = 6164Å)$, $\gamma(\lambda = 6334Å)$, and $\eta(\lambda = 6404Å)$.

The inset shows a 10 times enlarged part of the spectrum between 1800 and 2600 Å. It shows the excited neutral monatomic carbon $C_1^*$ emission lines denoted by the spectroscopic notation as the CI line emissions from the $3^1P_1$ level at $\lambda = 1931Å$, and $\lambda = 2478Å$. As the discharge current reduces to 100 and 50 mA in Figs. 1(b) and 1(c), respectively, the Ne line intensities reduce accordingly. The NIST Atomic Database [8] is used for the assignment of the Ne and C emission lines. The electron temperature $T_e$ and density $n_e$ of the Ne plasma are calculated from the relative intensities of the NeI and NeII emission lines. $T_e$ is calculated from the Boltzmann equation [9,10] by using the $\lambda = 3520$ and $\lambda = 5852Å$ lines of NeI; both lines result from the deexcitations of upper levels (4p and 3p) to the same level 3s[1/2]. A value of $T_e \leq 9000\pm1000K$ is obtained in the pressure range 0.1–0.6 mbar and for the range of $i_{dis}$= 50– 150 mA. The emission lines NeII $\lambda$ =and NeI $\lambda$ =5852Å are used in the Saha equation to estimate the electron density $n_e \approx 10^{12}$ cm$^{-3}$. In Fig. 2 the densities of the Ne I $3p'[1/2]$ level with the excitation energy $E_{exc}$= 18.72 eV and the NeII $^2D_0$ level with $E_{exc}$= 31.12 eV, are plotted as a function of the discharge current $i_{dis}$ between 50 and 200 mA.

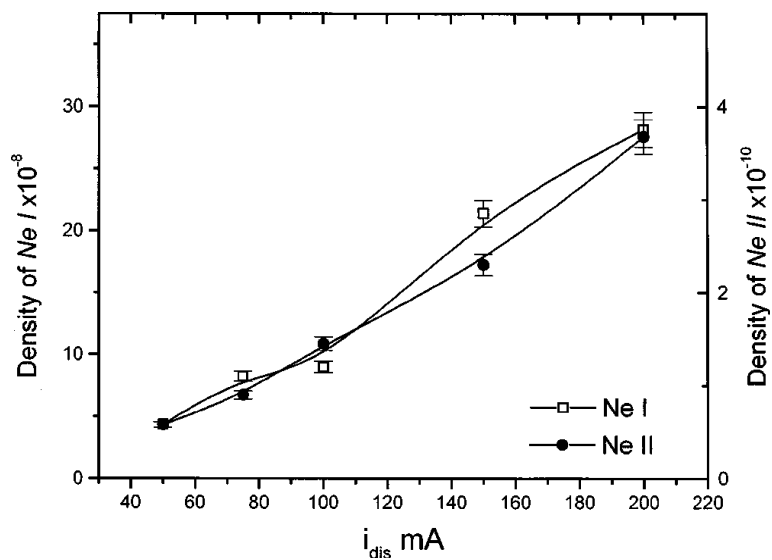

FIG. 2. Relative densities of Ne I and Ne II calculated from the spectra shown in Fig. 1. The Ne II line data are the average of the two lines at $\lambda = 3713$ and $\lambda = 3727$ Å, for the Ne II $^3P$– $^2D_0$ transition.

The number densities $N_m$ of the upper level $m$ for the transition to the lower $n$ is obtained by using $N_m = I_{mn}/(h\nu_{mn} A_{mn})$, where $I_{mn}$ is the intensity, $h\nu_{mn}$ the photon energy and $A_{mn}$ the Einstein transition probabilities of the respective emission lines. The spontaneous emissions of the three lines $\lambda = 5882$, $\lambda = 6030$, and $\lambda = 6164$Å shown as $\beta$, $\delta$, and $\phi$ in Figs. 1(a)–1(c) are all from the excited level $3p'[1/2]$ of NeI.

The average level densities along with those of the NeII $^2D_0$ level are plotted in Fig. 2. The level densities show similar rising trends with $i_{dis}$ for both the species. The average ratio of the ionized neon levels NeII $^2D_0$ to the excited atomic Ne I $3p'[1/2]$ levels is about $(1.3\pm0.3) \times 10^{-3}$ throughout the discharge range. The Boltzmann equation provides an estimate of the number densities of NeII from an existing density of Ne I in the $3p'[1/2]$ level, which implies that NeII/NeI $\approx 10^{-20}$ for $kT_e \approx 0.8$ eV. Therefore, the much higher ratio of the observed NeII fraction ($\approx 10^{-3}$) in the discharge cannot be due to the thermal electron ionization in the plasma, rather it is due to the high energy electrons that are emitted from the cathode and accelerated to 300–500 eV energies in the cathode fall region.

Each of these electrons ionizes ~10–15 metastable NeI atoms. The estimation of the ionized fraction (Ne$^+$ or NeII) of the source gas is essential for the kinetic sputtering and also for the secondary electron emission from the cathode. Both these phenomena are equally important for the generation of carbonaceous plasma and for the provision of the high energy electrons (300–500 eV) that are vital for the higher ionization cross sections of Ne.

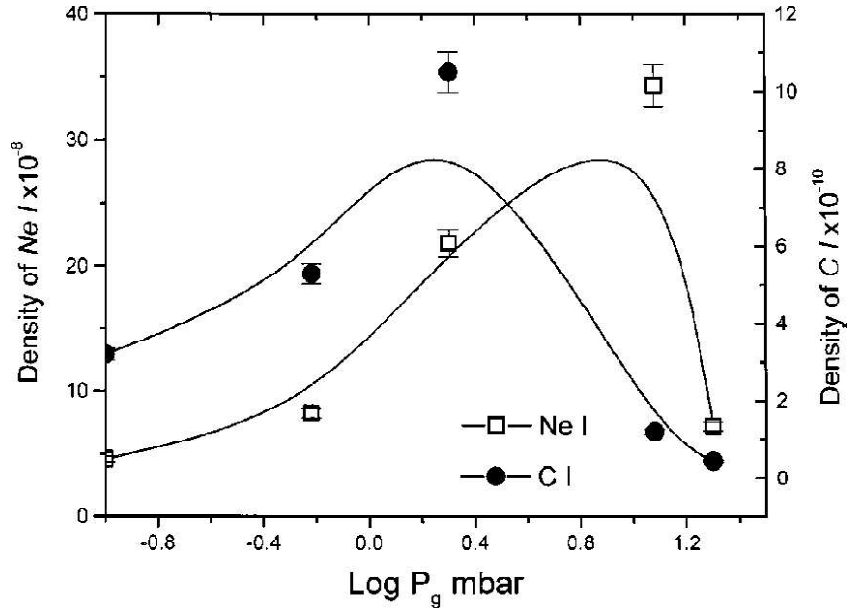

FIG. 3. Comparison of the excited level densities of the two discharge species C and Ne are shown as a function of the source gas pressure $\log P_g$.

Figure 3 shows the relative densities of the C and Ne excited levels $^1P_1$ and $3p'[1/2]$, respectively, plotted as a function of the source gas pressure $\log P_g$ between 0.1 and 20 mbar. The discharge current is $i_{dis} = 75$mA in all these experiments. The spline fit to the data points shows a maximum at $P_g$= 1.5 mbar for CI, and that of NeI is around 10 mbar. As both of these excited atoms decay to their respective metastable levels, the densities are indicators of the population densities as a function of the Ne pressures. The peaks for the two species are between 1 and 10 mbar.

The contributions of the two mechanisms of emission of C from the cathode can be compared to ascertain their relative contributions in the regeneration of the soot.

(1) Kinetic sputtering generates linear collision cascades by the incident ions which on interaction with the surface can emit the surface constituents as sputtered particles into the plasma. SRIM2000 [11] provides the sputtering yield =0.12±0.02 $C_1$ atoms/ion for $Ne_1$ ion with 500eV energy, incident on graphite.

(2) Potential sputtering of the sooted surface can take place upon the interaction of the Ne metastable atoms with 16.7 eV. Either an individual $_{C1}$ or a whole cluster $C_m$($m$>2) of many atoms adsorbed on the sooted cathode with binding energies $E(C_m) < E(Ne^*)$, can be ejected with the interaction of Ne* atoms [12]. These clusters can further go through the process of disintegration into smaller units or fragment into $m \times C_1$ atoms. The C atoms can be easily excited to a metastable level $^1S_0$ or $^1D_2$ by the thermal electrons in the plasma. At ~ 9000K, less than 0.7% remain in the ground state $^3P_{0,1,2}$ while 0.93 of all C atoms are in $^1D_2$ level while about 0.03 are in $^1S_0$ but only $10^{-4}$ in $^1P_1$ level. The corresponding CII/CI ratio is ~$10^{-6}$. The experimentally observed ratio is CI/NeII=4.5 ±30% in the pressure range 0.1–1 mbar and is relatively insensitive to the variations of the discharge current $i_{dis}$.

Similarly, the corresponding ratio is CI/NeI ≈5×$10^{-3}$. Using the kinetic sputtering yield data of SRIM2000 [11] for the ratio C1/NeII, we get ~0.12 $C_1$ atoms sputtered per NeII ion in the ground state $^3P_{0,1,2}$. This number density further reduces by a factor of $10^{-4}$ for exciting it to the CI $^1P_1$ level. Thus the kinetic sputtering will provide about $10^{-5}$ CI excited atoms in the $^1P_1$ level. This is much smaller than the observed relative density CI ≈ 4.5×NeII. The potential sputtering on the other hand, can also release between 0.5 and 1 CI per NeI [12]. This would yield (5 – 10)×$10^{-5}$ excited CI atoms for each NeI metastable atom, as the sputtering is occurring from a sooted cathode that is covered with many monolayers of a loose agglomeration of clusters. These clusters contain, on the average, 50–100 $C_1$ atoms [6]. On the inclusion of the potential sputtering of the clusters the estimates are within a factor of 3–4 of the observed value.

Therefore, we conclude that the observed high number density of CI is due to the potential sputtering of the carbon clusters adsorbed on the sooted surface by the NeI metastable atoms. This is our proposed dominant soot regenerating mechanism.